\documentclass[reprint, sort&compress]{elsarticle}
\usepackage{lineno}
\usepackage{graphicx}% Include figure files
\usepackage{dcolumn}% Align table columns on decimal point
\usepackage{bm}% bold math
\usepackage{upgreek}
\usepackage{hyperref}
\usepackage{ams math}
\usepackage{multirow}
\usepackage{soul}
\usepackage[margin=1.3in]{geometry}
\usepackage{bm}
\let\oldequation\equation
\let\oldendequation\endequation

\renewenvironment{equation}
  {\linenomathNonumbers\oldequation}
  {\oldendequation\endlinenomath}

\usepackage{mdframed} % Framing content

\usepackage{multicol} % Multiple columns environment

\usepackage{nomencl} % Nomenclature package

\makenomenclature

\usepackage{hyperref}
\usepackage{soul,color,xcolor}
\sethlcolor{yellow}

\soulregister{\cite}7
\soulregister{\citep}7
\soulregister{\citet}7
\soulregister{\ref}7
\soulregister{\pageref}7
\usepackage{etoolbox}
\renewcommand\nomgroup[1]{%
  \item[\bfseries
  \ifstrequal{#1}{G}{Greek symbols}{%
  \ifstrequal{#1}{S}{Subscripts}{}}%
]}

\journal{Journal of Colloid and Interface Science}

\begin{document}

\begin{frontmatter}

\title{Coalescence of immiscible droplets in liquid environments}

\author[SKLE]{Huadan Xu}
\author[SKLE,PES]{Tianyou Wang}
\author[SKLE,PES]{Zhizhao Che\corref{cor1}}
\cortext[cor1]{Corresponding author.}
\ead{chezhizhao@tju.edu.cn}
\address[SKLE]{State Key Laboratory of Engines, Tianjin University, Tianjin, 300350, China.}
\address[PES]{National Industry-Education Platform of Energy Storage, Tianjin University, Tianjin, 300350, China}

\begin{abstract}
\emph{Hypothesis}: Droplet coalescence process is important in many applications and has been studied extensively when two droplets are surrounded by gas. However, the coalescence dynamics would be different when the two droplets are surrounded by an external viscous liquid. The coalescence of immiscible droplets in liquids has not been explored.

\noindent \emph{Experiments}: In the present research, the coalescence of two immiscible droplets in low- and high-viscosity liquids is investigated and compared with their miscible counterparts experimentally. The coalescence dynamics is investigated via high-speed imaging, and theoretical models are proposed to analyze the growth of the liquid bridge.

\noindent \emph{Findings}: We find that, the liquid bridge $r$ evolves differently due to the constraint from the triple line in the bridge region, which follows $r\propto {{t}^{{2}/{3}}}$ for low-viscosity surroundings. While for high-viscosity surroundings, the liquid bridge grows at a constant velocity ${{u}_{r}}$ which varies with the surrounding viscosity ${{\mu }_{s}}$ as ${{u}_{r}}\propto {{\mu }_{s}}^{{1}/{2}}$. In the later stage of the bridge growth, the bridge evolution again merges with the well-established power-law regime $r\propto {{t}^{{1}/{2}}}$, being either in low or high-viscosity liquids. Moreover, a new inertia-viscous-capillary timescale is proposed, which unifies the combined influence of inertia, viscous, and capillary forces on the evolution of the liquid bridge in liquid environments, highlighting the joint role of inertia and viscous resistance in the coalescence process.
\end{abstract}

\begin{keyword}
\texttt {
Droplet coalescence \sep
Immiscible droplets \sep
Liquid bridge \sep
Liquid environment \sep
Fluid viscosity \sep
Inertia-viscous-capillary timescale
}
\end{keyword}
\end{frontmatter}

\begin{table}[]
\centering
\resizebox{\textwidth}{!}{%
\begin{tabular}{|llll|}
\hline
\multicolumn{2}{|l}{\textbf{Nomenclature}}&&\\
${Bo}$                  &Bond number                            &\multicolumn{2}{l|}{\textbf{Greek letters}}\\
$\overset{.}{{{{E}_{d}}}}$ &viscous dissipation                 &${\Delta A}$       &liquid bridge surface area\\
$\overset{.}{{{{E}_{\sigma}}}}$ &change rate of the interfacial energy &$\Delta {{E}_{k}}$ &kinetic energy\\
${{F}_{\nu }}$          &viscous stress                         &$\Delta {{E}_{s}}$ &interfacial energy \\
${{F}_{i }}$            &inertia stress                         &$\Delta p$     &capillary pressure \\
g                       &gravitational acceleration             &$\Delta {\rho }$ &density difference between droplet and surrounding liquid\\
${{l}_{\eta }}$         &characteristic length scale of ILV regime 	&${{\theta }_{12}}$ &contact angle of the oil and FC interface\\
${{l}_{ci}}$            &self-similar length scale                                                                              &$\kappa $          &liquid bridge curvature \\
${{Oh}_{tp}}$           &Ohnesorge number                       &${{\mu }_{d}}$  &viscosity of droplet \\
${r}$                   &liquid bridge radius                   &${{\mu }_{s}}$     &viscosity of surrounding liquid \\
${{r}^{*}}$             &nondimensional droplet radius          &${{\rho }_{1}}$    &density of oil droplet\\
${{R}_{1}}$             &oil droplet radius                     &${{\rho }_{2}}$    &density of FC droplet\\
${{R}_{2}}$             &FC droplet radius                      &${{\rho }_{d}}$    &density of droplet\\
${{R}_{e}}$             &effective radius of the two droplets   &${{\rho }_{s}}$    &density of surrounding liquid\\
${t}$                   &time                                   &${{\rho }_{e}}$    &equivalent density\\
${{t}^{*}}$             &rescaled time                          &${{\sigma }}$      &interfacial tension\\
${{t}_{h}}$             &bridge growth time                     &${{\sigma }_{12}}$ &interfacial tension between oil and FC  \\
${{t}_{\zeta }}$        &inertia-viscous-capillary timescale    &${{\sigma }_{1s}}$ &interfacial tension between oil and surrounding liquid \\
${{t}_{\eta }}$         &characteristic timescale of ILV regime &${{\sigma }_{2s}}$ &interfacial tension between FC and surrounding liquid \\
${{t}_{\mu}}$           &viscous timescale                      &${{\sigma }_{e}}$  &net equivalent interfacial tension \\
${{t}_{\mu }}^{*}$      &rescaled viscous timescale             &${{\tau }}_{\mathrm{out}}$ &fitted timescale  \\
${{t}_{\sigma }}$       &inertial timescale                     &\multicolumn{2}{l|}{\textbf{Abbreviations}}\\
${{t}_{\sigma }}^{*}$   &rescaled inertia timescale             &GS                 &glycerol solution \\
${{u}_{r}}$             &initial bridge expansion velocity      &IVC                &inertia-viscous-capillary \\
                        &                                       &ILV                &inertially limited viscous \\
                        &                                       &LED                &light-emitting diode \\
\hline
\end{tabular}%
}
\end{table}

\section{Introduction}\label{sec:1}
Droplet coalescence is a common phenomenon in many aspects of our life. A detailed and deeper understanding of coalescence is important in numerous situations, such as cloud formation \cite{Grabowski2013RainDroplet,denys2022lagrangian}, printing \cite{ihnen12}, and coating \cite{majumder10}. The coalescence of droplets in air has gained substantial attention since the pioneering work of Reynolds \cite{reynolds1881floating} a century ago. However, given the great importance of droplet coalescence in liquid environments in wide-ranging multiphase applications, i.e., emulsification in food manufacturing \cite{bera21, lobo03} and the biochemical reaction in bioscience \cite{thiam13}, rising attention is being paid to give a deeper understanding of this dynamics.

When two droplets come into contact, a thin film of the surrounding phase is formed between them. The thin liquid film will drain out gradually until it reaches a critical thickness, and the van der Waals force causes the thin film to eventually rupture \cite{jones78}. Following the rupture of the film, it has been shown that a liquid bridge is formed connecting the two coalescing droplets \cite{thoroddsen05speed, thoroddsen07, yao05}, which then expands over time. The bridge evolution dynamics has been extensively studied both experimentally and theoretically and remain to be an active area of research.

For two droplets coalescing in air, it can be approximate to the limit where the surrounding fluid has negligible viscosity and density, and the bridge growth is assumed to be dependent on the droplet properties (e.g., droplet radius, viscosity, surface tension, and density) \cite{Paulsen2013}. The growth of the liquid bridge, driven by the capillary force, could either be opposed by the viscous force or inertia. Therefore, depending on the dominant resistance, two main regimes have been classified, which are characterized by two different scaling laws. For highly viscous droplets (for example, higher than 82 mPa$\cdot$s), the growth of the liquid bridge radius $r$ was found to follow a linear growth $r\propto t$ as the viscous force dominates over inertia and balances the capillary force \cite{paulsen11}. In contrast, for the case of approximately inviscid droplets, as the inertia dominates over viscous forces, the bridge radius evolves as $r\propto {{t}^{{1}/{2}}}$, as has been reported in many studies \cite{duchemin03, eggers99, thoroddsen05speed}.

For droplets coalescing in liquid environments, many studies have tried to disentangle the influences of the surrounding liquids on the coalescence process. First, the increased viscosity of the surroundings was found to delay the drainage substantially \cite{janssen11}, which was also observed in the experiment of Paulsen et al.\ \cite{paulsen14}. In the experiments of Aryafar et al.\ \cite{aryafar08}, by setting the viscosity of the surrounding liquid to be substantially large, it was discovered that the larger viscosity of the droplet and the surrounding fluids determines the coalescence speed. Jose et al.\ \cite{jose17} later, by developing a viscous function, quantified the cooperative yet asymmetrical influence of the liquid viscosities during the spreading and coalescence of droplets while immersed in liquids. In the theoretical work of Mitra et al.\ \cite{mitra15}, a much slower bridge growth was observed at a later time of bridge growth. Further, in studying the coalescing of surfactant-laden droplets in liquid environments, Nowak et al.\ \cite{nowak16} discovered that the high viscosity of the outer fluids could weaken the influence of surfactants by retarding the interfacial flow. Recently, with the multiple benefits of electric fields being used for water-oil separation in crude oil refining, interest has also been raised in examining charged droplets coalescing in liquid surroundings. The viscosity of the surrounding liquid was found to reduce the chance of coalescence \cite{sadeghi18}, which was believed to originate from an increased resistance from the thin liquid film between the two droplets. While for two oppositely charged droplets coalescing in a viscous oil, due to the viscous stress from the surrounding oil, the increased oil viscosity increases the critical cone angle in the liquid bridge for the transition from coalescence to pinch-off \cite{chen19}.

Yet, in the above-mentioned works regarding droplet coalescence in liquid environments, attention has been predominantly paid to the coalescence of two miscible droplets. However, recently, synthetic chemistry and biology have seen a wide application of immiscible droplets such as the crystallization of membrane proteins \cite{chen07} and interfacial catalytic reactions \cite{crossley10}. On those occasions, droplets of different components that are not miscible with each other are dispensed into another immiscible liquid. Motivated by strong practical interests, one question would arise: what would the coalescence of immiscible droplets in liquids be like when they come into contact? When two immiscible droplets contact, there forms an immiscible interface separating the two droplets during and after the coalescence process \cite{zhang20}. This immiscible interface not only prevents mass exchange \cite{baumgartner19} but also influences energy conversion \cite{bernard20, xu2022bridge}. For the existing studies directly or indirectly involving the interaction of immiscible droplets in liquids, some focus on the very early stage when the liquid film drains out \cite{choi14}, while others focus predominantly on the dynamics in a much larger timescale, for example, multi-component droplets in confined geometries \cite{chen07, liu23, zhang15} and collision and engulfment of impacting immiscible droplets in a liquid environment \cite{ebadi22, liu22}. Based on the above-mentioned previous literature and to the best of our knowledge, the short-time dynamics characterizing the growth of the liquid bridge during the coalescence of immiscible droplets in liquid environments have never been examined. With the objective of providing theoretical insights into droplet coalescence and achieving precise control in relevant applications, here, we present an experimental study of two immiscible droplets coalescing in liquid environments. The two immiscible droplets are kept to be approximately inviscid while the viscosity of the surrounding liquids is varied in a wide range, as we aim to investigate how the viscosity of the outer liquids influences the coalescence process. For comparison, experiments with miscible counterparts are also conducted. The bridge evolution process is studied qualitatively first, including their morphology and bridge growth time. For a more quantitative study, the influence of the outer fluids on the growth of the liquid bridge of immiscible droplets is discussed and compared with that of miscible droplets. In addition, a new inertia-viscous-capillary scaling was proposed to incorporate the cooperative effects of inertia and viscous resistance during the liquid bridge evolution.

\begin{table}[]
\centering
\caption{Physical properties of the liquids used in this study at 20$^\circ$C. The glycerol solutions (GS) were prepared by mixing glycerol and NaCl aqueous solution (2.8 wt\%), and the percentage values in the parenthesis represent the mass fraction of glycerol. The values of densities are from Ref.~\cite{takamura12}; the values of viscosities are from Ref.~\cite{sheely32}; the values of interfacial tensions were measured in the laboratory using the pendent droplet method \cite{BERRY2015226}.}
\label{tab:1}
\scalebox{0.8}{
\begin{tabular}{lllll}
\hline
Liquid &
  \begin{tabular}[c]{@{}l@{}}Density\\ $\rho$ (kg/m$^3$)\end{tabular} &
  \begin{tabular}[c]{@{}l@{}}Viscosity\\ $\mu$ (mPa$\cdot$s)\end{tabular} &
  \begin{tabular}[c]{@{}l@{}}Interfacial   tension \\ with oil phase\\ $\sigma$ (mN/m)\end{tabular} &
  \begin{tabular}[c]{@{}l@{}}Interfacial   tension \\ with FC phase\\ $\sigma$ (mN/m)\end{tabular} \\
\hline
Silicone   oil (2 cSt) & 880    & 2.2   & /                  & $ 5.04   \pm 0.9$ \\
FC 40                  & 1855   & 1.76  & $5.04   \pm 0.1  $ & /                  \\
GS (0\%)               & 1026   & 1.0   & $40.1   \pm 1.1  $ & $ 52.1   \pm 0.7$ \\
GS (20\%)              & 1075   & 1.76  & $32.1   \pm 0.7 $ & $ 45.4   \pm 0.9$ \\
GS (60\%)              & 1181   & 10.9  & $27.3   \pm 0.8 $ & $ 35.6   \pm 0.7$ \\
GS (70\%)              & 1187.5 & 22.9  & $26.1   \pm 0.5 $ & $ 34.3   \pm 0.2$ \\
GS (80\%)              & 1226   & 62.0  & $26.8   \pm 0.3 $ & $ 34.1   \pm 0.6$ \\
GS (85\%)              & 1232   & 112.9 & $25.1   \pm 0.7 $ & $ 33.1   \pm 0.9$ \\
GS (86\%)              & 1238.5 & 129.6 & $24.9   \pm 1.5 $ & $ 32.8   \pm 0.8$ \\
GS (88\%)              & 1243   & 174.5 & $24.3   \pm 1.2 $ & $ 31.4   \pm 1.5$ \\
GS (89\%)              & 1248   & 201.4 & $24.2   \pm 1.1 $ & $ 30.9   \pm 1.2$ \\
GS (90\%)              & 1251   & 234.6 & $23.9   \pm 1.3 $ & $ 30.9   \pm 1.1$ \\
\hline
\end{tabular}}
\end{table}

\section{Materials and method}\label{sec:2}
\subsection{Liquids used}\label{sec:21}
In the experiments, to consider the influence of the viscosity of the external liquids on the coalescence of immiscible droplets, several groups of aqueous glycerol solutions were prepared to vary the viscosity of the surrounding liquids. NaCl was dissolved into the solution to increase the refractive index of the surroundings to facilitate imaging. The addition of NaCl (with 2.8\% by weight) into the aqueous glycerol solutions was tested to induce slight changes in the liquid properties, namely interfacial tension, density, and viscosity (less than 3\%) \cite{takamura12}. Then, silicone oil droplets (kinetic viscosity: 2 cSt) and FC-40 droplets were used, as they are not only immiscible with each other \cite{WACKER20111891}, but also have a much lower viscosity compared to the surrounding liquid. Hereafter they are abbreviated as ‘oil droplet’ and ‘FC droplet’, respectively. In the present three-phase immiscible liquid system dealing with droplet coalescence, the two droplet phases and the surrounding phase are labeled as 1 (for the oil droplet), 2 (for the FC droplet), and s (for the surrounding liquid), which are used in the subscripts of physical parameters for the corresponding three phases, respectively. The interfacial tension between the droplet phase and the surrounding phase was determined using the pendent droplet method \cite{BERRY2015226}. The densities are from Ref.~\cite{takamura12} and the viscosities are from the measurements of Ref.~\cite{sheely32}. The measurements of the fluid properties and the experiments of droplet coalescence were conducted at a temperature of 20 $^\circ$C. The properties of the liquids used in this study are listed in Table \ref{tab:1}.

\subsection{Experimental setup}\label{sec:22}
\begin{figure}
  \centering
  \includegraphics[scale=0.7]{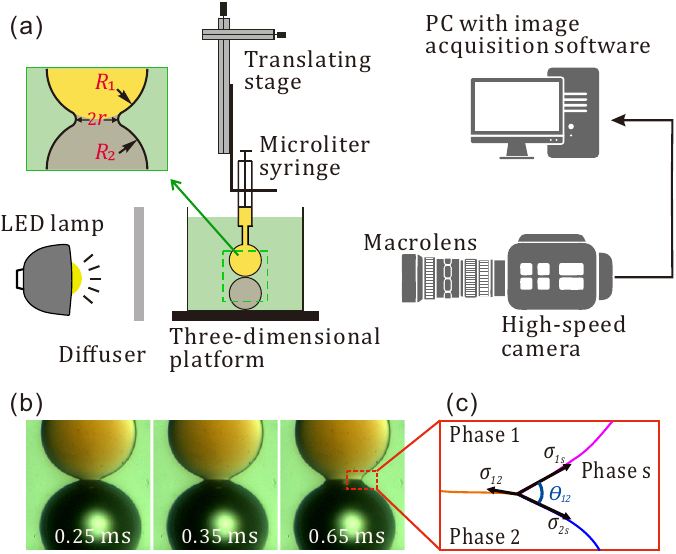}
  \caption{(a) Schematic diagram of the experimental setup. (b) Typical images of the coalescing droplets (the oil droplet was colored with oil red O). (c) Sketch of the three-liquid interfaces in the bridge region, where phases ``1'', ``2'', and ``s'' represent the oil phase, the FC liquid phase, and the surrounding phase, respectively.}\label{fig:01}
\end{figure}

The experimental setup is schematically shown in Figure \ref{fig:01}a. The two droplets were immersed in a deep transparent glass tank which is 5 cm $\times$ 5 cm, being large enough for the wall effects to be ignored. The two droplets were aligned vertically before the coalescence by adjusting their position using three-dimensional translating stages (XYZ displacement table: 120 mm $\times$ 80 mm $\times$ 50 mm). The upper droplet (Silicone oil droplet) was controlled by a microliter syringe of 1.0 $\upmu$l capacity which aims to accurately control the droplet size and avoid the additional momentum while introducing the droplet into the liquids. The lower droplet (FC droplet) was pre-deposited on a hydrophobic substrate on the bottom surface of the glass tank. The hydrophobic surface was obtained by soaking soda-lime glass in the hydrophobic solution (MesoPhobic-2000; MesoBioSystem), which was characterized by the contact angle of about $150 \pm 5 ^\circ$ (obtained through ImageJ software of 15 measurements) for the FC droplet in different glycerol solutions in our experiments. (It is noted that this hydrophobic surface could degrade when used for a long time, but here this hydrophobic surface was used to keep the free surfaces of the sessile FC droplet before the contact to be spherical enough and away from the substrate to minimize the effect of the substrate. Therefore, the experimental results would not be affected as long as the FC droplet has a contact angle larger than $90^\circ$ on the substrate). Further, to avoid the potential interference from the electrostatic effect after the droplet is dispensed from a micropipette \cite{Choi2013}, we first ground the upper droplets with a ground wire connecting the tip of the syringe and then ground the aqueous glycerol solution in the glass tank.

The experiments were first achieved by bringing the upper oil droplet and the lower FC droplet in each other's close vicinity, where the distance between the center of the two droplets is equal to the sum of the droplet radius. Then, the coalescence was initiated after the drainage of the liquid film between the two droplets. The drainage time increases with the viscosity of the surrounding liquids. Sometimes, when the viscosity of the surrounding liquid is high (e.g., higher than 100 mPa$\cdot$s), the two droplets could remain stable without coalescence for minutes or even longer. This can be attributed to the drainage process of the liquid film, which becomes very slow due to the increased viscous resistance. Such a phenomenon of non-coalescence has been reported by related works \cite{couder05, neitzel02}. Here, we do not attempt to investigate the detailed drainage process but rather focus on the liquid bridge expansion which happens after the collapse of the liquid film between the two immiscible droplets.
\subsection{Visualization and recording}\label{sec:23}

To capture details of the microscopic dynamics, which is the bridge evolution in the present experiment, a high-speed camera (Phantom V1612) was employed. Videos were recorded at a frame rate of 120,000 frames per second. Moreover, a microscopic lens (Navitar Zoom 6000) was adopted to accomplish close-up views which allows for the highest resolution of 3.58 µm/pixel. A glass diffuser was mounted in front of a light-emitting diode (LED) light source to achieve a uniform background for lighting. The subsequent measurement of the bridge radius ${r}$ started from the instant when we could detect a first increase in the bridge radius as adopted in many experiments \cite{eddi13, yao05}. Further, to ensure we capture the early dynamics of the coalescence process and the whole process of liquid bridge growth, we started the recording quite a long time before the contact of the two droplets, and the data with an initial radius larger than 8\% of the initial droplet radius was rejected in the analysis. Typical images of the coalescence of immiscible droplets in liquid environments are shown in Figure \ref{fig:01}b. An edge-detection algorithm in MATLAB was then performed to track the evolution of the bridge radius in post-processing the high-speed images.

\subsection{Parameters of droplets}\label{sec:24}
\begin{figure}
  \centering
  \includegraphics[scale=0.7]{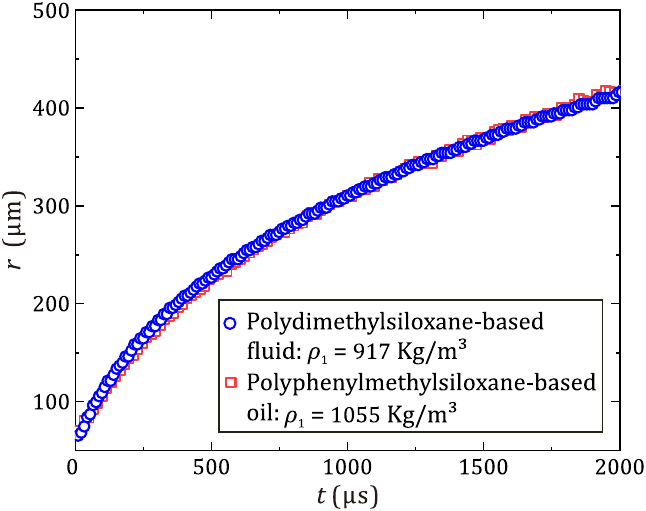}
  \caption{Growth of the liquid bridge during the coalescence of oil droplets with FC droplets in water (${{\mu }_{s}}$ = 1 mPa$\cdot$s). The different oil droplets have similar interfacial tension (around 40 mN/m) and viscosity (around 50 mPa$\cdot$s) but different relative densities to the surrounding liquid. The polydimethylsiloxane-based fluid (i.e., silicone oil used in this study) has a smaller density than the surrounding water, while the polyphenylmethylsiloxane-based oil has a larger density than the surrounding water.}\label{fig:02}
\end{figure}

In the experiments, the droplet diameter was set to be around 1 mm, which is smaller than their capillary length ${{l}_{c}}=\sqrt{{\sigma }/{\rho g}}$ (which is around 1.6 mm for the FC droplet and 2.1 mm for the silicone oil droplet). The influence of gravity on the droplet coalescence can be characterized by the Bond number $Bo={\Delta \rho g{{R}^{2}}}/{\sigma }$, where $\Delta \rho $ is the density difference between the droplets and the surrounding liquid, $R$ is the droplet radius and $\sigma $ is the interfacial tension between the droplet and the surrounding liquid. For the oil droplet and the FC droplet, the Bond numbers are around 0.01 and 0.02, respectively. Furthermore, since the silicone oil droplet used here has a lower density than the surrounding liquids, we also tested polyphenylmethylsiloxane-based oil for comparison, which has a density larger than water. The bridge growth does not show obvious differences between each other as can be seen in Figure \ref{fig:02}. This further demonstrates the weak influence of the buoyancy force. Then, during the coalescence process, the main forces involved are capillary, viscous, and inertia. The capillary force serves as the driving force of the liquid bridge motion while inertia and/or viscosity are the resistance. Depending on the different dominant forces, there would be different timescales to characterize the growth of the liquid bridge, and the detailed dimensional analysis is discussed and delineated in Section \ref{sec:32}.

\section{Results and discussion}\label{sec:3}
\subsection{Coalescence process in liquid environments}\label{sec:31}

\begin{figure}
  \centering
  \includegraphics[scale=0.7]{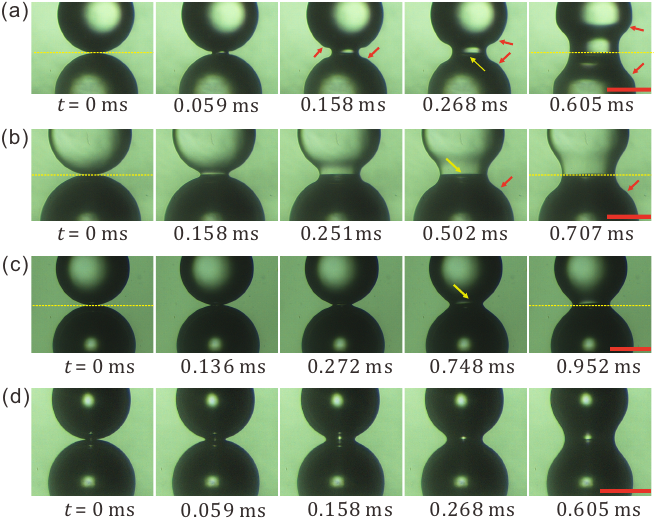}
  \caption{(a--c) Time sequences of the coalescence process of immiscible droplets in surrounding liquids of different viscosities: (a) 1 mPa$\cdot$s, (b) 10.9 mPa$\cdot$s, and (c) 62.0 mPa$\cdot$s. The upper droplet is 2-cSt oil and the lower droplet is FC-40. (d) Time sequences of the coalescence process of miscible droplets in surrounding liquid of 62.0 mPa$\cdot$s, where the two droplets are both FC-40. The scale bars are 500 $\upmu$m. The frame just before the contact of the two droplets is defined as $t = 0$. (Video clips for these processes are available as Supplementary Material Movies 1--4).}\label{fig:03}
\end{figure}

Figure \ref{fig:03} shows the liquid bridge growth process during the coalescence of two immiscible droplets. Following the first contact of the two droplets, the formation of the meniscus in the bridge area generates high capillary pressure which then drives the rapid expansion of the liquid bridge. One unique feature is shown in the bridge region, where a water-oil-FC triple line (Figure \ref{fig:03}a) or a GS-oil-FC triple line (Figure \ref{fig:03}b) is produced after the two immiscible droplets contact (as is evident in Figure \ref{fig:03}a-b, also in Supplementary Movies 1 and 2); this feature is absent in the case of miscible droplets. The dynamical evolution of this triple line is governed by inertia and viscous hydrodynamics and is subjected to the constraint of contact angles in the bridge area. The growing liquid bridge adopts a sharper shape compared to that of miscible droplets coalescing at the same surrounding viscosity (see Figure \ref{fig:03}d and Supplementary Movie 4) and forms a wedge-like region around the triple line. The sharp region corresponds to a high capillary pressure in the bridge area which indicates a high viscous stress around the triple line. Another distinctive feature, compared to miscible droplets, is the unique immiscible interface of two immiscible droplets as indicated by the yellow arrows in Figure \ref{fig:03}. The thin yellow dotted lines in the first and last panels in Figure \ref{fig:03}a-c denote the initial contact position of the two droplets (Here, we only mark the first and last snapshots of each group to have a better observation of the interface from the other three snapshots). The position of this immiscible interface is observed to be approximately unchanged with time. This suggests a remarkable difference from the coalescence of miscible droplets with interfacial tension differences in liquids, where the penetration of the high-interfacial tension droplet and a strong jet flow were observed inside the droplet of lower interfacial tension \cite{luo19}. This restricted motion of the immiscible interface in the axial direction can be attributed to the high capillary pressure in the wedge-like region around the triple line \cite{cuttle21}.

In addition to the morphology changes near the liquid bridge shown in Figure \ref{fig:03}a, obvious capillary waves can be observed on the surface of both droplets (see also Supplementary Movie 1), when the surrounding phase has a low viscosity during the coalescence. Similar capillary waves also occur in situations such as droplet coalescence in air \cite{thoroddsen07}, or bubble coalescence in liquids \cite{oratis23, thoroddsen05}. When the viscosity of the surrounding liquid is increased to ${{\mu }_{s}}$ = 10.96 mPa$\cdot$s, the bridge dynamics is qualitatively similar to that for the two immiscible droplets coalescing in water (${{\mu }_{s}}$ = 1 mPa$\cdot$s), except that the magnitude of the capillary wave is greatly weakened as shown in Figure \ref{fig:03}b and Supplementary Movie 2. For an even higher viscosity of the surrounding phase (${{\mu }_{s}}$ = 62.0 mPa$\cdot$s, see Figure \ref{fig:03}c and Supplementary Movie 3), the growth of the liquid bridge is comparatively slower (as evident from the timestamps below the snapshots in Figure \ref{fig:03}c). This decreased motion of the liquid bridge can be understood from a reversed physical situation of a retracting rim: increased viscosity increases the viscous resistance of retraction and, in this case, slows down the bridge expansion.

It should be noted that in the present configuration of immiscible droplet coalescence, the sum of the interfacial tension between the oil droplet and the surrounding liquid (${{\sigma }_{1s}}$) and the interfacial tension of the oil droplet and FC droplet (${{\sigma }_{12}}$) is smaller than the interfacial tension between the FC droplet and the surrounding liquid (${{\sigma }_{2s}}$), i.e., ${{\sigma }_{1s}}+{{\sigma }_{12}}<{{\sigma }_{2s}}$. This would create a net interfacial tension force along the FC droplet surface, and therefore, a Marangoni flow, which has been commonly seen in the coalescence of miscible droplets with different surface tensions \cite{hack21, thoroddsen07}. However, in the present situation of immiscible droplet coalescence, the expected Marangoni flow would be additionally subjected to the constraint of the contact angles near the triple line formed after the contact of the two immiscible droplets \cite{cuttle21}. For example, an unsmooth interface in the bridge area is observed with a wedge-like shape around the triple line (see Figure \ref{fig:01}b for example). This shows a clear difference from the coalescence of miscible droplets (see Figure \ref{fig:03}d) where the surface is smooth. Further, such an interface cusp suggests that the Marangoni flow is restrained, with no oil layer being drawn onto the FC droplet surface, at least in the early stage of the bridge growth. This can be attributed to the constraint from the triple line in the bridge area where three contact angles evolve dynamically. Therefore, the Marangoni flow would not show up in the early stage of the bridge growth but would further develop and play a role in the encapsulation of the higher-interfacial droplet in a longer timescale \cite{chen17, planchette10}. However, this is beyond the scope of the present study.

\subsection{Dynamics of liquid bridge growth}\label{sec:32}
To better understand the bridge growth dynamics of coalescence, we first take a closer look at the different forces involved in the coalescence process: capillary, viscous, and inertial forces. The capillary force drives the motion of the liquid bridge while inertia and/or viscosity resist the motion. Therefore, two timescales can be derived according to dimensional analysis, which is the inertial timescale, ${{t}_{\sigma }}$, and the viscous timescale, ${{t}_{\mu }}$. For the coalescence of immiscible droplets in liquid environments, balancing the inertial stress and capillary forces, the inertial timescale is given as:
\begin{equation}\label{eq:01}
  {{t}_{\sigma }}=\sqrt{\frac{2{{\rho }_{e}}{{R}_{e}}^{3}}{{{\sigma }_{e}}}},
\end{equation}
then, if balancing the viscous stress with capillary forces, the viscous timescale can be obtained as:
\begin{equation}\label{eq:02}
  {{t}_{\mu }}=\frac{2{{\mu }_{s}}{{R}_{e}}}{{{\sigma }_{e}}},
\end{equation}
which are based on the net equivalent interfacial tension ${{\sigma }_{e}}={{\sigma }_{1s}}+{{\sigma }_{2s}}-{{\sigma }_{12}}$ (see Figure \ref{fig:01}c). Half of this net equivalent interfacial tension ${{\sigma }_{e}}/2$ is used in Eqs.\ (\ref{eq:01}) and (\ref{eq:02}) to keep consistent with the configuration of the two same droplets coalescence, in which ${{\sigma }_{12}}=0$, ${{\sigma }_{1s}}={{\sigma }_{2s}}={{\sigma }_{e}}/2$. In addition, ${{\rho }_{e}}={{\rho }_{1}}+{{\rho }_{2}}-{{\rho }_{s}}$ is used as the equivalent density, and ${{R}_{e}}={2}/{({1}/{{{R}_{1}}}+{1}/{{{R}_{2}}})}$ is the effective radius of the two droplets (where $R_1$ and $R_2$ are the radii of the oil and FC droplets, respectively). In contrast, for the coalescence of miscible droplets (two FC droplets coalescing in liquid environments), the two corresponding characteristic timescales are given as:
\begin{equation}\label{eq:03}
  {{t}_{\sigma }}=\sqrt{\frac{{{\rho }_{e}}{{R}_{e}}^{3}}{{{\sigma }_{2s}}}},
\end{equation}
\begin{equation}\label{eq:04}
  {{t}_{\mu }}=\frac{{{\mu }_{s}}{{R}_{e}}}{{{\sigma }_{2s}}},
\end{equation}
where the interfacial tension between the FC droplet and the surrounding liquid ${{\sigma }_{2s}}$ is used in the definition.

\begin{figure}
  \centering
  \includegraphics[scale=0.7]{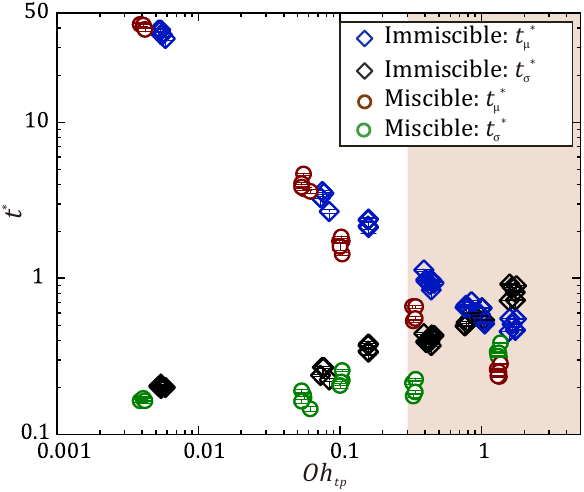}
  \caption{Rescaled time of the bridge growth (${{t}_{\sigma }}^{*}={{{t}_{h}}}/{{{t}_{\sigma }}}$ and ${{t}_{\mu }}^{*}={{{t}_{h}}}/{{{t}_{\mu }}}$) when the bridge radius grows to half of the effective radius of the coalescing droplets as a function of the Ohnesorge number ${{Oh}_{tp}}$. The time ${{t}_{h}}$ has been rescaled by the inertia timescale ${{t}_{\sigma }}$ and the viscous timescale ${{t}_{\mu }}$, where ${{t}_{\sigma }}=\sqrt{{2{{\rho }_{e}}{{R}_{e}}^{3}}/{{{\sigma }_{e}}}}$ and ${{t}_{\mu }}={2{{\mu }_{s}}{{R}_{e}}}/{{{\sigma }_{e}}}$ for the coalescence of immiscible droplets; and ${{t}_{\sigma }}=\sqrt{{{{\rho }_{e}}{{R}_{e}}^{3}}/{{{\sigma }_{2s}}}}$ and ${{t}_{\mu }}={{{\mu }_{s}}{{R}_{e}}}/{{{\sigma }_{2s}}}$ for the coalescence of miscible droplets. The error bars (smaller than the size of the symbol) indicate the deviation brought by the measurement of ${{t}_{h}}$.}\label{fig:04}
\end{figure}

By comparing the inertial and viscous timescales, an Ohnesorge number can be obtained to characterize the coalescence dynamics. For the coalescence of immiscible droplets in liquids, the Ohnesorge number is defined as:
\begin{equation}\label{eq:05}
  {{Oh}_{tp}}=\frac{{{\mu }_{s}}}{\sqrt{{{\rho }_{e}}{{R}_{e}}{{{\sigma }_{e}}}/{2}}}.
\end{equation}
For the coalescence of miscible droplets in liquids, because ${{\sigma }_{1s}}={{\sigma }_{2s}}$ and ${{\sigma }_{12}}=0$, the Ohnesorge number becomes
\begin{equation}\label{eq:06}
  {{Oh}_{tp}}=\frac{{{\mu }_{s}}}{\sqrt{{{\rho }_{e}}{{R}_{e}}{{\sigma }_{2s}}}}.
\end{equation}
To first look into the bridge growth process, we measure the time required for the bridge radius to grow to half of the initial effective radius of the two droplets, ${{t}_{h}}$. Then, to further quantitatively investigate the influence of the viscous resistance from the surroundings, the time of the bridge growth ${{t}_{h}}$ is normalized by the inertial timescale ${{t}_{\sigma }}$ and the viscous timescale ${{t}_{\mu }}$. Figure \ref{fig:04} shows the rescaled time of bridge growth as a function of the defined Ohnesorge number ${{Oh}_{tp}}$. The rescaled time (${{t}_{\sigma }}^{*}$ and ${{t}_{\mu }}^{*}$) shows different trends with the increase of ${{Oh}_{tp}}$. ${{t}_{\mu }}^{*}$ decreases with ${{Oh}_{tp}}$ and is much greater than ${{t}_{\mu }}^{*}$ at small ${{Oh}_{tp}}$ (e.g., ${{Oh}_{tp}}< 0.1$). The two rescaled times become comparable when ${{Oh}_{tp}} > 0.2$, indicating that viscous resistance begins to exert a nonnegligible influence on the bridge growth, which corresponds to ${{\mu }_{s}}> 62$ mPa$\cdot$s in the present experiments.

In the experiments, the two droplets are both kept to be approximately inviscid while the surrounding liquid viscosity gradually increases to be as high as ${{\mu }_{s}}= 234.6$ mPa$\cdot$s. According to the results in Figure \ref{fig:04}, we can see different trends for low-viscosity surrounding liquids and high-viscosity surrounding liquids. Hence, we next investigate the liquid bridge growth during the coalescence of immiscible droplets from two aspects: for low-viscosity surrounding liquid (${{\mu }_{s}} < 62$ mPa$\cdot$s in our case, in Section \ref{sec:321}) and for high-viscosity surrounding liquid (${{\mu }_{s}} > 62$ mPa$\cdot$s in our case, in Section \ref{sec:322}).

Noted that, at early times, the expansion of the liquid bridge was found always to start at the inertially limited viscous (ILV) regime, which is dominated by the inner droplet liquids, even for highly viscous surrounding liquids \cite{paulsen14}. This is because the relative effect of the viscous force is not only affected by the fluid viscosity but also affected by the length scale. The ILV regime happens at a characteristic length and time scale \cite{yao05viscous} of
\begin{equation}\label{eq:07}
  {{l}_{\eta }}=\frac{{{\mu }_{d}}^{2}}{{{\rho }_{d}}\sigma },
\end{equation}
\begin{equation}\label{eq:08}
  {{t}_{\eta }}=\frac{{{\mu }_{d}}^{3}}{{{\rho }_{d}}{{\sigma }^{2}}},
\end{equation}
where ${{\mu }_{d}}$ is the droplet viscosity, ${{\rho }_{d}}$ is the droplet density, and $\sigma $ is the interfacial tension for the situation of droplet coalescence in liquids. In practice, ${{l}_{\eta }}$ is often very small and ${{t}_{\eta }}$ is very short, which makes it hard to observe this viscous regime in experiments \cite{aarts05}. For example, for the low-viscosity droplets used in the experiments, we can find ${{l}_{\eta }}=3\times {{10}^{-8}}$ m and ${{t}_{\eta }}=3\times {{10}^{-9}}$ s. Thus, when $r\gg {{l}_{\eta }}$, the effect of surrounding liquids will be dominant for a large part of the bridge evolution in liquids. Beyond the ILV regime, for the influence of the external liquids, it was identified that, for two bubbles coalescing in liquids \cite{paulsen14, Anthony2017}, variation of the liquid viscosity leads to the bubble coalescence in two regimes, which are viscous regime for high viscous external liquids and inertial regime for low viscous external liquids. The liquid bridge growth was found to follow the $r\propto {{{t}/{t}_{c}}^{{1}/{2}}}$ law for both regimes, while dominated by different timescales. That is, ${{t}_{c }}={{{\mu }}{{R}}}/{{{\sigma }}}$ for the viscous regime; and ${{t}_{c }}=\sqrt{{{{\rho }}{R}^{3}}/{{{\sigma }}}}$  for the inertia regime, where ${\mu }$ and ${\rho }$ are the viscosity and density of the external liquids, respectively \cite{paulsen14}. In this study, our focus is on the regime where the surrounding liquids influence the bridge dynamics at time and length scales much larger than that of the ILV regime. The ILV regime is then not considered in this study. Here, following a similar consideration, we next investigate the coalescence of immiscible droplets in low- and high-viscosity liquids, respectively.

\subsubsection{Immiscible droplet coalescence in low-viscosity surrounding liquids}\label{sec:321}
\begin{figure}
  \centering
  \includegraphics[scale=0.65]{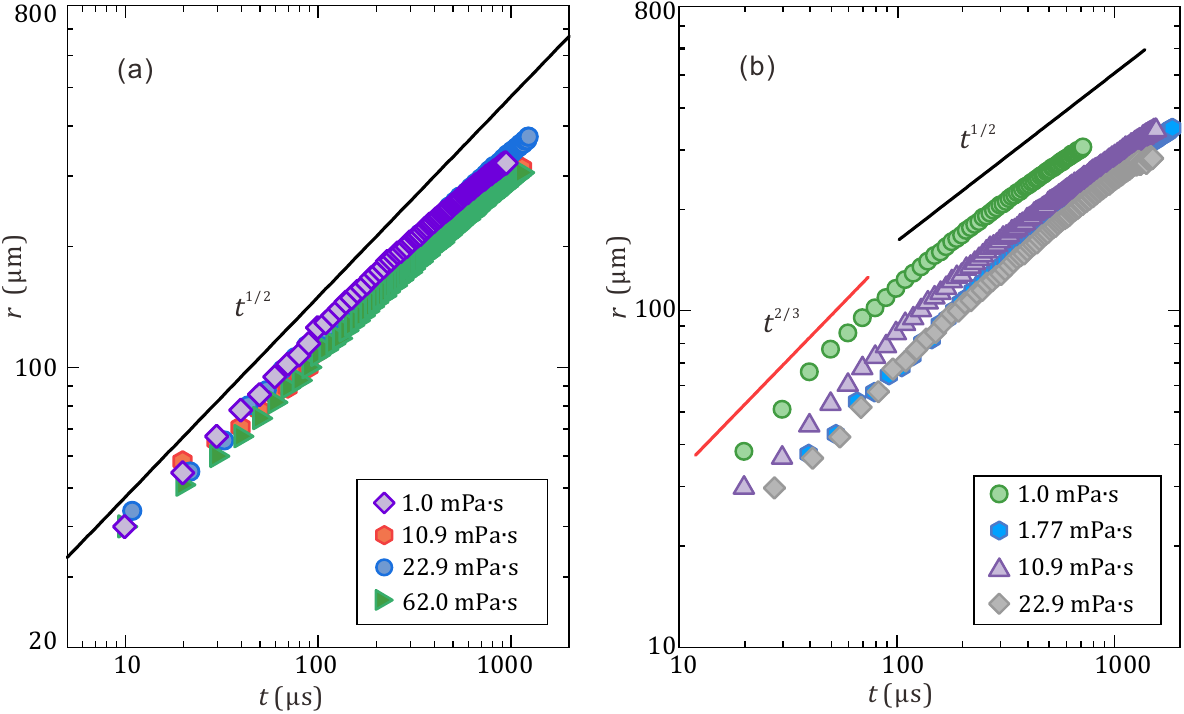}
  \caption{Bridge evolution in low-viscosity surrounding liquids for the coalescence of (a) miscible droplets, and (b) immiscible droplets.}\label{fig:05}
\end{figure}

To further investigate the liquid bridge dynamics of immiscible droplet coalescence in liquid environments, we first look into the coalescence in low-viscosity surrounding liquids, which, in the present experiment, corresponds to the proposed ${{Oh}_{tp}}$ being much smaller than 0.2 (${{\mu }_{s}}<60$ mPa$\cdot$s). Figure \ref{fig:05} shows the bridge radius $r$ as a function of time after contact, which reveals the power-law growth of the liquid bridge.

To be clearer, we first analyze the growth of the liquid bridge from the miscible configuration. We can see that for the miscible droplets (two FC droplets in low-viscosity liquids, see Figure \ref{fig:05}a), the bridge follows the 1/2 power law growth, $r\propto {{t}^{{1}/{2}}}$. This 1/2 power-law growth can be derived based on the balance between capillary pressure and inertia stress, which are the two dominant stresses when the surrounding liquids are approximately inviscid. This regime is called the inertia-dominated regime, as has been defined in the coalescence of miscible droplets \cite{thoroddsen07}. The capillary pressure $\sigma \kappa $ serves as the driving force, where the curvature $\kappa $ is obtained from the small length scale ${{{r}^{2}}}/{{{R}_{e}}}$ as given by Duchemin et al.\ \cite{duchemin03}. The capillary pressure $\Delta p$ is then expressed as:
\begin{equation}\label{eq:09}
  \Delta p\sim{{\sigma }_{2s}}\frac{{{R}_{e}}}{{{r}^{2}}},
\end{equation}
and the inertia stress of the growing bridge is given as:
\begin{equation}\label{eq:10}
  {{F}_{i}}\sim{{\rho }_{e}}{{u}_{r}}^{2},
\end{equation}
where ${{u}_{r}}={dr}/{dt}$ is the expansion velocity of the bridge radius. By balancing the two dominant forces in Eqs.\ (\ref{eq:09}) and (\ref{eq:10}), we can obtain a differential equation for the expansion dynamics of the liquid bridge:
\begin{equation}\label{eq:11}
  {{\rho }_{e}}{{\left( \frac{dr}{dt} \right)}^{2}}\sim {{\sigma }_{2s}}\frac{{{R}_{e}}}{{{r}^{2}}},
\end{equation}
Through the integration of Eq.\ (\ref{eq:11}), we can have
\begin{equation}\label{eq:12}
    r\sim {{\left(\frac{{{\sigma }_{2s}}{{R}_{e}}}{{{\rho }_{e}}}\right)}^{{1}/{4}}}{{t}^{{1}/{2}}}.
\end{equation}
The obtained 1/2 power law growth in Eq.\ (\ref{eq:12}) describes well the scaling laws of the bridge dynamics of miscible droplet coalescence, as shown in Figure \ref{fig:05}a.

Next, we move on to the immiscible droplet configuration. For the liquid bridge in the coalescence of immiscible droplets, in the early time of droplet coalescence, there exists an immiscible interface between the two immiscible droplets forming the triple line in the bridge region (as illustrated in Section \ref{sec:31}). Further, the extra immiscible interfacial tension (${{\sigma }_{12}}$) would serve as a retraction force and influence the flow in the bridge region, which is absent in the miscible counterpart. Therefore, the bridge dynamics would be expected to show different behaviors. After the two immiscible droplets contact, the bridge area resembles a wedge geometry due to the constraint of the triple line. Such initial configuration of liquid bridge evolution is similar to related problems of surface-tension-driven flow (for example, flow in wedges and cones) \cite{billingham99, keller00}, which does not have an external geometrical length scale. The surface-tension-driven flow in the wedge area was found to be limited by the self-similar capillary waves generated at the contact line \cite{keller00}, where the flow field varies over an independent length scale, ${{l}_{ci}}\sim{{({\sigma {{t}^{2}}}/{\rho })}^{{1}/{3}}}$, as noted by Keller and Miksis \cite{keller83}.

Inspired by these ideas, we then consider the bridge evolution from the perspective of energy conservation, where the change of the interfacial energy balances with the kinetic energy. First, the magnitude of the velocity can be estimated by the bridge expansion velocity $u\sim{dr}/{dt}$. The change of the surface area during the coalescence can be estimated as $\Delta A\sim{{l}_{ci}}r$, with which the variation of interfacial energy can be obtained:
\begin{equation}\label{eq:13}
  \Delta {{E}_{s}}\sim{{\sigma }_{e}}\Delta A\sim {{\sigma }_{e}}{{l}_{ci}}r,
\end{equation}
and the kinetic energy is given by:
\begin{equation}\label{eq:14}
  \Delta {{E}_{k}}\sim{{\rho }_{e}}{{u}^{2}}\Delta V\sim {{\rho }_{e}}{{\left( \frac{dr}{dt} \right)}^{2}}{{l}_{ci}}^{3}.
\end{equation}

In the case of low-viscosity liquids, when the viscous dissipation is comparatively small, the conservation between the interfacial energy and kinetic energy gives:
\begin{equation}\label{eq:15}
  {{\rho }_{e}}{{\left( \frac{dr}{dt} \right)}^{2}}{{l}_{ci}}^{3}\sim{{\sigma }_{e}}{{l}_{ci}}r.
\end{equation}
By setting ${{l}_{ci}}\sim {{({{{{{\sigma }_{e}}{{t}^{2}}}/{\rho }}_{e}})}^{{1}/{3}}}$, we can have:
\begin{equation}\label{eq:16}
  \frac{dr}{dt}\sim {{\left( \frac{{{\sigma }_{e}}}{{{\rho }_{e}}} \right)}^{{1}/{6}}}{{t}^{{-2}/{3}}}{{r}^{{1}/{2}}}.
\end{equation}
Then, by separation of variables and integration, we can obtain:
\begin{equation}\label{eq:17}
  r\sim {{\left( \frac{{{\sigma }_{e}}}{{{\rho }_{e}}} \right)}^{{1}/{3}}}{{t}^{{2}/{3}}},
\end{equation}
which gives a 2/3 power law relationship for the bridge growth, i.e., $r\propto {{t}^{{2}/{3}}}$. As shown in Figure \ref{fig:05}b, the scaling in Eq.\ (\ref{eq:17}) agrees well with the experimental data.

This unique self-similar regime of immiscible droplet coalescence, $r\propto {{t}^{{2}/{3}}}$, reflects the influence of the triple line on the liquid bridge. During the initial expansion when the contact angle of the oil and FC interface, ${{\theta }_{12}}$ (see the schematic in Figure \ref{fig:01}c), is much smaller than 180$^\circ$, the restriction of the triple-line in the bridge region destroys the quadratic relationship of the curvature, i.e., $\kappa \propto {R}/{{{r}^{2}}}$ \cite{eggers99}. In other words, the constraint of the triple line results in a flow similar to the dynamics of “merging wedges” \cite{keller83}, which is also characterized by similar 2/3 self-similar flow dynamics. While in the later stage of the liquid bridge growth, as the bridge radius increases and ${{\theta }_{12}}$ evolves close to 180$^\circ$, the influence of the triple line constraint gradually weakens. The curvature of the bridge again conforms to the quadratic shape of the spherical droplets, satisfying $\kappa \propto {R}/{{{r}^{2}}}$, and that is why the square-root behavior ($r\propto {{t}^{{1}/{2}}}$) is then recovered.

\subsubsection{Immiscible droplet coalescence in high-viscosity surrounding liquids}\label{sec:322}

For high-viscosity surrounding liquids, the viscous resistance is much more significant than the inertial effects. In the present experiment, this corresponds to the situation when the proposed ${{Oh}_{tp}}$ is much larger than 0.2 (${{\mu }_{s}}>60$ mPa$\cdot$s). In such a scenario, the dynamics of the liquid bridge is governed by viscous stress from the surrounding phase ${{F}_{\nu }}$ and capillary pressure $\Delta p$ \cite{paulsen11}. Following the same logical order as that in Section \ref{sec:321}, we again start our analysis with the coalescence of miscible droplets and then proceed to the immiscible counterpart.

\begin{figure}
  \centering
  \includegraphics[width=0.8\columnwidth]{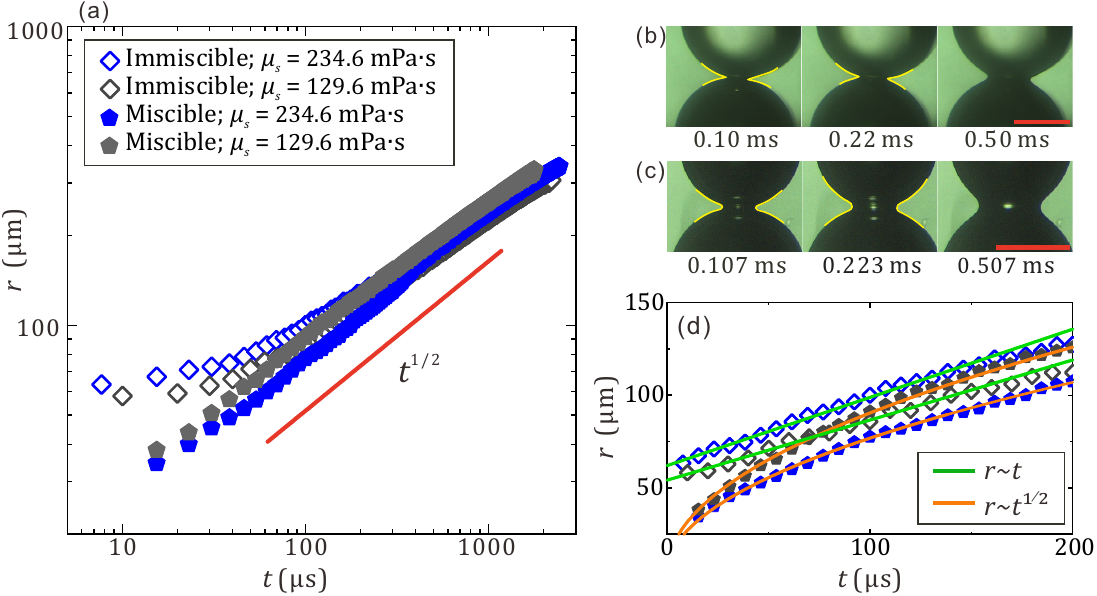}
  \caption{(a) Temporal evolution of the bridge radius for high-viscosity surrounding liquids. Typical snapshots of (b) miscible droplets and (c) immiscible droplets coalescing in viscous surrounding liquids of 129.6 mPa$\cdot$s. The yellow outlines in (b) denote the wedge-shaped interface in the bridge region, while the yellow outlines in (c) denote the smooth shape in the bridge region. The scale bars represent 0.5 mm. (d) At early time, the bridge radius of immiscible droplets varies linearly with time; while the growth of the bridge radius of miscible droplets follows the 1/2 power-law.}\label{fig:06}
\end{figure}

Since the two droplets are approximately inviscid, the dominant viscous stress is in the radial direction with the length scale of $r$, which then gives:
\begin{equation}\label{eq:18}
  {{F}_{\nu }}={{\mu }_{s}}\frac{{{u}_{r}}}{r}.
\end{equation}
Moreover, by setting ${{u}_{r}}={dr}/{dt}$ and balancing the viscous stress with the capillary pressure expressed in Eq.\ (\ref{eq:09}), we can then obtain a differential equation describing the evolution of the liquid bridge:
\begin{equation}\label{eq:19}
  {{\mu }_{s}}{{r}^{-1}}\frac{dr}{dt}\sim {{\sigma }_{2s}}\frac{{{R}_{e}}}{{{r}^{2}}},
\end{equation}
further rearrangement and integration of Eq.\ (\ref{eq:19}) can lead to
\begin{equation}\label{eq:20}
  r\sim {{\left( \frac{{{\sigma }_{2s}}{{R}_{e}}}{{{\mu }_{s}}} \right)}^{{1}/{2}}}{{t}^{{1}/{2}}}.
\end{equation}
This prediction is further supported by the coalescence data of miscible droplets in Figure \ref{fig:06}a, which shows that the liquid bridge $r(t)$ evolves with ${{t}^{{1}/{2}}}$.

One may wonder about the same 1/2 power-law scaling of the liquid bridge of miscible droplets in highly viscous surroundings as that in low-viscosity surroundings (i.e., the same 1/2 slope in the log-log plot in Figure \ref{fig:05}a and Figure \ref{fig:06}a). The rationale behind this same exponent comes from the same stress scale, where the length scale $r$ is the dominant gradient of stress. Therefore, when inserting $r(t)$ into the stress scale of Eqs.\ (\ref{eq:11}) and (\ref{eq:19}), we can obtain the same 1/2 power law of each regime (but with different timescales).

When it comes to the coalescence of immiscible droplets in highly viscous liquids, the results of the bridge growth are shown to be different from that of miscible droplets. The difference can first be seen in Figure \ref{fig:06}a: in the early time (e.g., $t<100 $ $\upmu$s), bridge growth deviates from the 1/2 power law growth. Further, when we zoom in on the growth of the liquid bridge in this early stage and replot the bridge radius in the linear coordinate system, a linear regime is observed, i.e., the liquid bridge of immiscible droplets grows linearly with time (see Figure \ref{fig:06}c). It should be noted that this linear growth fails to show in the log-log plot in Figure \ref{fig:06}a, as the bridge does not start from zero due to the drainage of the liquid film \cite{chan11}. To further explore the liquid bridge dynamics of immiscible droplet coalescence, we, therefore, quantify this bridge velocity next and see how it varies with the surrounding viscosities.

\begin{figure}
  \centering
  \includegraphics[scale=0.7]{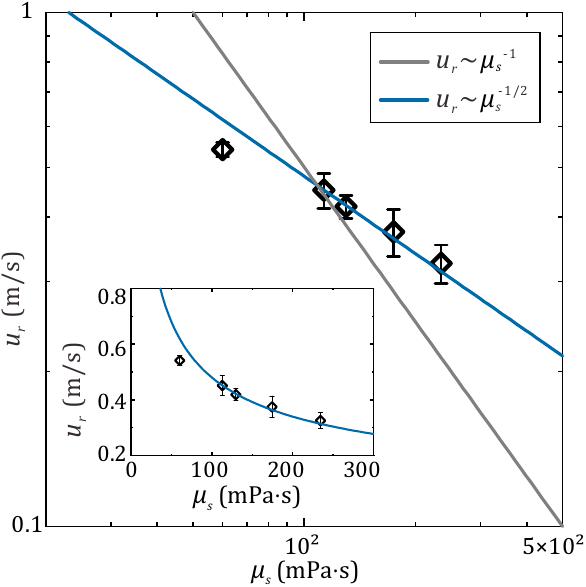}
  \caption{Variation of initial bridge expansion velocity (${{u}_{r}}$) with the dynamic viscosity of the surrounding liquids (${{\mu }_{s}}$) for the coalescence of immiscible droplets in liquid environments. The graph is shown in a log-log scale with the inset plotted in a linear-linear scale. The discrete data points are the mean of experimental measurements from five independent experiments, and the error bars represent the standard deviation.}\label{fig:07}
\end{figure}

This velocity of bridge growth is expected to be set by the capillary velocity ${\sigma }/{\mu }$, as is common in situations of viscous-dominated capillary flows \cite{aarts05, aryafar08, eri10}. Therefore, in this case, the bridge velocity is estimated to be proportional to the inverse of the viscosity ($\sim {1}/{{{\mu }_{s}}}$). However, the result of the initial expansion velocity of the bridge ${{u}_{r}}$ shows that, although ${{u}_{r}}$ decreases with the viscosity of the surrounding liquids, the dependence is much weaker than the expected $\sim{1}/{{{\mu }_{s}}}$. It can be seen from Figure \ref{fig:07} that the expansion velocity is only reduced by a factor of around 0.6 when the surrounding liquid viscosity is increased by a factor of 4.

The rationale behind this weak dependence on ${1}/{{{\mu }_{s}}}$ may go to the high dissipation rate in the wedge-liked region near the triple line. Moreover, as shown in Figure \ref{fig:06}b, the bridge area of immiscible droplets shows a much sharper interface, suggesting high viscous stress in the local area near the triple line. Similar wedge flow has also been observed in situations such as bubble receding in liquids and film retraction on liquids and was reported to have a high viscous dissipation rate in the bridge region near the triple line \cite{Reyssat06, sanjay22}. We then consider the balance between the viscous dissipation and the released interfacial energy. In the spirit of the flow in the wedge-like area near the triple line \cite{sanjay22}, the viscous dissipation is given as:
\begin{equation}\label{eq:21}
  \overset{.}{{{{E}_{d}}}}\sim {{Oh}_{tp}}^{{1}/{2}}{{{\sigma }}_{{e}}}^{2}{{{\mu }}_{s}}^{-1}\left( 2{\pi r}(t) \right).
\end{equation}
The change rate of the interfacial energy is:
\begin{equation}\label{eq:22}
  \overset{.}{{{{E}_{\sigma }}}}\sim{{F}_{\sigma }}\overset{.}{{r}}(t)\sim2{{{\sigma }}_{{e}}}(2{\pi r}(t)){{{u}}_{r}}.
\end{equation}
Equating Eqs.\ (\ref{eq:21}) and (\ref{eq:22}), we can first obtain that $u_r$ scales as:
\begin{equation}\label{eq:23}
  {{u}_{r}}\sim{{\mu }_{s}}^{{-1}/{2}}\Delta {{\sigma }^{{3}/{4}}}{{\rho }_{e}}^{{-1}/{4}}{{R}_{e}}^{{-1}/{4}},
\end{equation}
which is time-invariant. Moreover, according to Eq.\ (\ref{eq:23}), $u_r$ is found to scale with viscosity as ${{u}_{r}}\propto {{\mu }_{s}}^{{-1}/{2}}$, which gives a better fit for the relation between the initial bridge velocity ${{u}_{r}}$ and the viscosity of the surrounding liquid ${{\mu }_{s}}$ as shown in Figure \ref{fig:07}.

Shortly after the linear growth stage, with the further increase of the liquid bridge, its growth is observed to merge with the 1/2 power law scaling as shown in Figure \ref{fig:06}a. This is because in the later stage, as the constraint from the contact angles around the triple line further relaxes, the viscous dissipation in the triple line area gradually decreases. Therefore, the viscous stress in the radial direction ($\sim r$) dominates, which supports the same 1/2 power law growth as that of the miscible droplets as discussed in the initial part of this section.

\subsubsection{ Inertia-viscous-capillary scale: a universal scaling for different viscosities of surrounding liquids}\label{sec:323}
In this part, we further discuss the 1/2 power-law growth for the surrounding liquids of both low and high viscosities. Returning to the results in Figure \ref{fig:04}, we can find that the two dimensionless timescales ${{t}_{\sigma }}^{*}$ and ${{t}_{\mu }}^{*}$ (which are rescaled by the characteristic inertial timescale ${{t}_{\sigma }}$ and viscous timescale ${{t}_{\mu }}$, respectively) are comparable with each other when ${{Oh}_{tp}}$ is larger than 0.2. Even when ${{Oh}_{tp}}$ is around 0.1 (which corresponds to ${{\mu }_{s}}$ = 22.9 mPa$\cdot$s), the two rescaled times seem to be close to the order of 1. This indicates a nonnegligible influence from either inertia or viscous resistance and the liquid bridge growth is predicted to be governed by an inertial-viscous-capillary balance. Here, it should be noted that, when further increasing the viscosity of the surrounding liquids, the difference between the two rescaled times would increase again if ${{t}_{\mu }}^{*}$ is significantly larger than ${{t}_{\sigma }}^{*}$. However, the increased viscosity of the surrounding liquids would make it hard for the intermediate liquid film between the two droplets to be drained out to initiate the coalescence. Therefore, the viscosity of the surrounding fluids in our experiment only goes as far as  234.6 mPa$\cdot$s.

Proceeding under this consideration, we first start with the analysis of the coalescence of miscible droplets since it is much easier to arrive at a clear force expression. By balancing the capillary force expressed in Eq.\ (\ref{eq:09}) with the inertia and viscous forces given in Eqs.\ (\ref{eq:10}) and (\ref{eq:18}), respectively, we can obtain a differential equation describing the expansion of the liquid bridge:
\begin{equation}\label{eq:24}
  2\frac{{{R}_{e}}{{\sigma }_{2s}}}{{{r}^{2}}}\sim {{\rho }_{e}}{{\left( \frac{dr}{dt} \right)}^{2}}+{{\mu }_{s}}\frac{dr}{dt}{{r}^{-1}}.
\end{equation}
By solving this equation, we can first obtain:
\begin{equation}\label{eq:25}
  \frac{dr}{dt}\sim \frac{-{{\mu }_{s}}+\sqrt{{{\mu }_{s}}^{2}+8{{\rho }_{e}}{{R}_{e}}{{\sigma }_{2s}}}}{2{{\rho }_{e}}}.
\end{equation}
Additionally, we can use the properties defined in the definition of ${{Oh}_{tp}}$ to non-dimensionalize Eq.\ (\ref{eq:25}) and have the solution through further integration:
\begin{equation}\label{eq:26}
  {{r}^{*}}\sim \frac{1}{2}{{\left(\sqrt{{{Oh}_{tp}}^{2}+8}-{{Oh}_{tp}}\right)}^{1/2}}\sqrt{{{t}^{*}}},
\end{equation}
where ${{r}^{*}}={r}/{{{R}_{e}}}$, ${{t}^{*}}={t}/{{{t}_{\sigma }}}$, and ${{t}_{\sigma }}$ is the same definition given in Eq.\ (\ref{eq:01}).

According to Eq.\ (\ref{eq:26}), we can define a new timescale that characterizes the combined resistance of inertia and viscous forces, henceforth referred to as an inertia-viscous-capillary (IVC) timescale:
\begin{equation}\label{eq:27}
  {{t}_{\zeta }}=4{{{t}_{\sigma }}}/{\left( \sqrt{{{Oh}_{tp}}^{2}+8}-{{Oh}_{tp}} \right)}.
\end{equation}
Then, Eq.\ (\ref{eq:26}) could be rewritten in a more compact form as:
\begin{equation}\label{eq:28}
  {{r}^{*}}\sim \sqrt{{t}/{{{t}_{\zeta }}}}.
\end{equation}

\begin{figure}
  \centering
  \includegraphics[width=0.8\columnwidth]{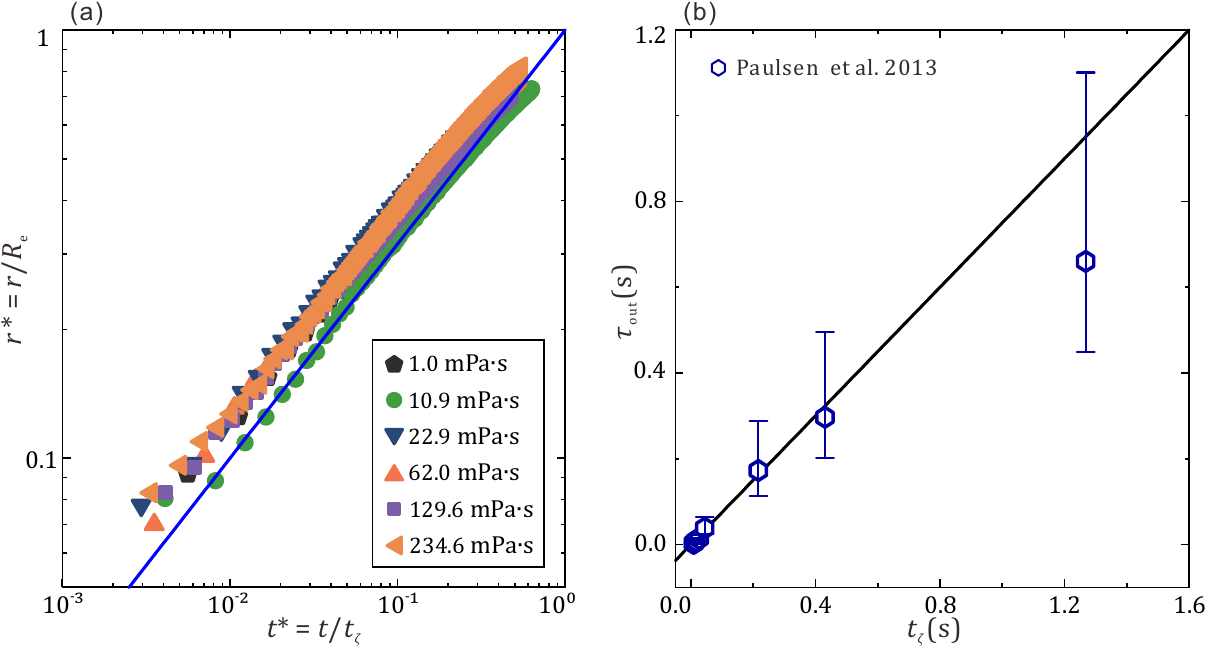}
  \caption{(a) Evolution of liquid bridge of miscible droplets in surrounding liquids, rescaled by initial effective droplet radius (${{R}_{e}}$) and the proposed IVC timescale (${{t}_{\zeta }}$) in Eq.\ (\ref{eq:27}). The surrounding-liquid viscosity varies from 1.0 to 234.6 mPa$\cdot$s. (b) Timescale ${{ {\tau }}_{\mathrm{out}}}$ in the work of Paulsen et al.\ \cite{paulsen14} (obtained by fitting) versus the proposed IVC timescale ${{t}_{\zeta }}$. The data follow ${{ {\tau }}_{\mathrm{out}}}=0.75{{t}_{\zeta }}$ with a wide viscosity range (0.49--2900 mPa$\cdot$s), indicating the proposed IVC timescale ${{t}_{\zeta }}$ can characterize the combined inertia and viscous resistances of surrounding liquids in the coalescence of approximately inviscid droplets.}\label{fig:08}
\end{figure}

To further test the newly proposed IVC timescale ${{t}_{\zeta }}$, the liquid bridge length and time are then rescaled with the effective droplet radius ${{R}_{e}}$ and ${{t}_{\zeta }}$. All the data are found to be well collapsed onto the curve of ${{r}^{*}}={t^*}^{1/2}$ (see the solid blue line in Figure \ref{fig:08}a). Moreover, the data in the experiment of Paulsen et al.\ \cite{paulsen14} are also compared, in which the value of ${{ {\tau }}_{\mathrm{out}}}$ was obtained through fitting. By plotting ${{ {\tau }}_{\mathrm{out}}}$ against ${{t}_{\zeta }}$ (see Figure \ref{fig:08}b), we can see a linear relationship ${{ {\tau }}_{\mathrm{out}}}=0.75{{t}_{\zeta }}$ within the wide viscosity range (0.49--2900 mPa$\cdot$s). This relationship between ${{ {\tau }}_{\mathrm{out}}}$ and ${{t}_{\zeta }}$ further supports our hypothesis that the proposed IVC timescale ${{t}_{\zeta }}$ successfully characterizes the combined inertia and viscous resistance of surrounding liquids in the coalescence of approximately inviscid droplets.

\begin{figure}
  \centering
  \includegraphics[scale=0.7]{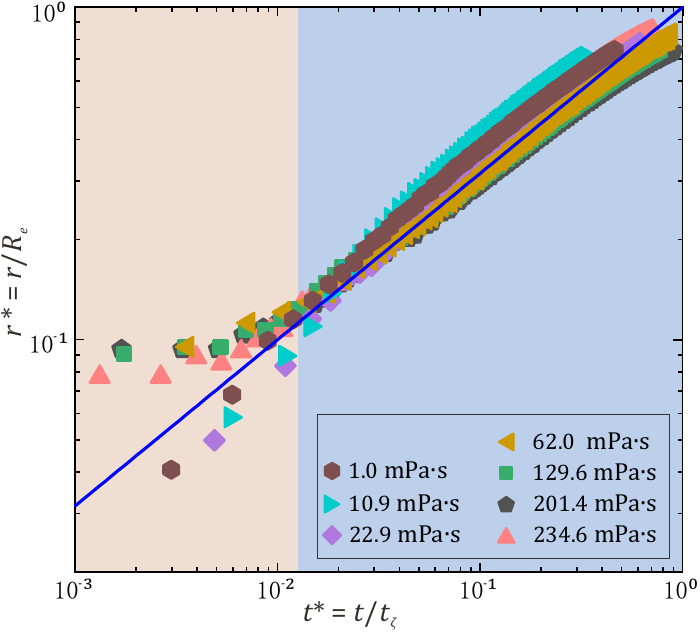}
  \caption{Evolution of liquid bridge of immiscible droplets in surrounding liquids, rescaled by initial effective droplet radius (${{R}_{e}}$) and the proposed IVC timescale (${{t}_{\zeta }}$). The surrounding-liquid viscosity varies from 1.0 to 234.6 mPa$\cdot$s. The blue-shaded region indicates the regime where the constraint from the triple line is weakened. The solid blue line is $r^*={t^*}^{1/2}$.}\label{fig:09}
\end{figure}

We next apply the proposed IVC timescale ${{t}_{\zeta }}$ to the bridge dynamics in the coalescence of immiscible droplets. By adopting the same scaling methods as that of the miscible droplet in Figure \ref{fig:08}a using the proposed IVC timescale ${{t}_{\zeta }}$, the data were found to be again successfully collapsed into a master curve (see the solid blue line $r^*={t^*}^{1/2}$ in Figure \ref{fig:09}), except the initial stage where the triple line constraint poses an extra influence to the bridge dynamics. Based on the above results, it can be obtained that the proposed IVC timescale ${{t}_{\zeta }}$ captures the cooperative influence of inertial and viscous resistance in the bridge evolution driven by capillary force, and can be applied to characterize the coalescence of both miscible and immiscible droplets. Moreover, the successful collapse of the data of immiscible droplets, being similar to that of miscible droplets, indicates that in the late stage where the constraint of the triple line weakens, the growing liquid bridge is controlled by the similar curvature relation and similar inertia and viscous resistance.

\section{Conclusions}\label{sec:4}
In the present work, we have studied the coalescence of immiscible droplets in liquid environments and compared it to the miscible counterparts. Due to the existence of the triple line, the liquid bridge evolution of immiscible droplets shows different dynamics. For low-viscosity surrounding phases, due to the constraint of the triple line in the early stage of bridge growth, the flow in the bridge area resembles the configuration of capillary-driven wedge flow. In this case, the growth of the liquid bridge is limited by the self-similar capillary waves generated at the triple line, the liquid bridge radius of immiscible droplets is found to first evolve as ${{t}^{{2}/{3}}}$. For the high-viscosity surrounding phase, the liquid bridge first grows at a constant speed. The speed is found to show a weaker dependence on the viscosity ${{u}_{r}}\propto {{\mu }_{s}}^{-1/2}$ than the expected relation ($\propto {{\mu }_{s}}^{-1}$) due to the high dissipation in the wedge region around the triple line. More importantly, compared to previous works \cite{yao05viscous, yokota11}, which mainly consider the resistance of viscous forces or inertia separately, we have proposed a new IVC scaling ${{t}_{\zeta }}$ considering the balance of both inertia and viscous resistances with the capillary force. By collapsing the liquid bridge data of both miscible and immiscible droplets in our experiment and comparing this IVC timescale ${{t}_{\zeta }}$ with the characteristic timescale obtained in previous work, we validate that the proposed ${{t}_{\zeta }}$ is a more suitable timescale for characterizing the evolution of liquid bridge in the liquid environment of varying viscosities.

This work further supplements the studies on droplet coalescence dynamics. In practical applications in biological and petrochemical industries, the presence of additives (such as surfactants \cite{baret12}, polymers \cite{beck-broichsitter21}, and microparticles \cite{tian19}) is common in many real situations and could bring different influences on mass transfer and efficient mixing/reaction for applications. Therefore, future work can focus on the coalescence behavior in the presence of additives to provide a deeper insight into the local hydrodynamics, which would help to increase the predictability of the reactor design and to improve the efficiency of related processes.

\section*{Acknowledgements}
This work is supported by the National Natural Science Foundation of China (Grant Nos.\ 52176083, 51920105010, and 51921004).

\section*{Appendix A.\ Supplementary material}
Supplementary material associated with this article can be found in the online version.

\bibliography{immiscibleCoalescenceInLiquid}

\begin{thebibliography}{10}
\expandafter\ifx\csname url\endcsname\relax
  \def\url#1{\texttt{#1}}\fi
\expandafter\ifx\csname urlprefix\endcsname\relax\def\urlprefix{URL }\fi
\expandafter\ifx\csname href\endcsname\relax
  \def\href#1#2{#2} \def\path#1{#1}\fi

\bibitem{Grabowski2013RainDroplet}
W.~W. Grabowski, L.-P. Wang, Growth of cloud droplets in a turbulent
  environment, Annu. Rev. Fluid Mech. 45 (2013) 293--324.
\newblock \href {https://doi.org/10.1146/annurev-fluid-011212-140750}
  {\path{doi:10.1146/annurev-fluid-011212-140750}}.

\bibitem{denys2022lagrangian}
M.~Denys, P.~Deuar, Z.~Che, P.~E. Theodorakis, A lagrangian particle-based
  numerical model for surfactant-laden droplets at macroscales, Phys. Fluids
  34~(9) (2022) 095126.
\newblock \href {https://doi.org/10.1063/5.0101930}
  {\path{doi:10.1063/5.0101930}}.

\bibitem{ihnen12}
A.~C. Ihnen, A.~M. Petrock, T.~Chou, B.~E. Fuchs, W.~Y. Lee, Organic
  nanocomposite structure tailored by controlling droplet coalescence during
  inkjet printing, ACS Appl. Mater. Interfaces 4~(9) (2012) 4691--4699.
\newblock \href {https://doi.org/10.1021/am301050n}
  {\path{doi:10.1021/am301050n}}.

\bibitem{majumder10}
M.~Majumder, C.~Rendall, M.~Li, N.~Behabtu, J.~A. Eukel, R.~H. Hauge, H.~K.
  Schmidt, M.~Pasquali, Insights into the physics of spray coating of {SWNT}
  films, Chem. Eng. Sci. 65~(6) (2010) 2000--2008.
\newblock \href {https://doi.org/10.1016/j.ces.2009.11.042}
  {\path{doi:10.1016/j.ces.2009.11.042}}.

\bibitem{reynolds1881floating}
O.~Reynolds, On the floating of drops on the surface of water depending only on
  the purity of the surface, Proc. Lit. Phil. Soc. Manchester 21~(1) (1881)
  413--414.

\bibitem{bera21}
B.~Bera, R.~Khazal, K.~Schroën, Coalescence dynamics in oil-in-water emulsions
  at elevated temperatures, Sci. Rep. 11~(1) (2021).
\newblock \href {https://doi.org/10.1038/s41598-021-89919-5}
  {\path{doi:10.1038/s41598-021-89919-5}}.

\bibitem{lobo03}
L.~Lobo, A.~Svereika, Coalescence during emulsification. 2. role of small
  molecule surfactants, J. Colloid Interface Sci. 261~(2) (2003) 498--507.
\newblock \href {https://doi.org/10.1016/S0021-9797(03)00069-9}
  {\path{doi:10.1016/S0021-9797(03)00069-9}}.

\bibitem{thiam13}
A.~R. Thiam, R.~V. Farese~Jr, T.~C. Walther, The biophysics and cell biology of
  lipid droplets, Nat. Rev. Mol. Cell Biol. 14~(12) (2013) 775--786.
\newblock \href {https://doi.org/10.1038/nrm3699} {\path{doi:10.1038/nrm3699}}.

\bibitem{jones78}
A.~Jones, S.~Wilson, The film drainage problem in droplet coalescence, J. Fluid
  Mech. 87~(2) (1978) 263--288.
\newblock \href {https://doi.org/10.1017/S0022112078001585}
  {\path{doi:10.1017/S0022112078001585}}.

\bibitem{thoroddsen05speed}
S.~T. Thoroddsen, K.~Takehara, T.~G. Etoh, The coalescence speed of a pendent
  and a sessile drop, J. Fluid Mech. 527 (2005) 85--114.
\newblock \href {https://doi.org/10.1017/s0022112004003076}
  {\path{doi:10.1017/s0022112004003076}}.

\bibitem{thoroddsen07}
S.~Thoroddsen, B.~Qian, T.~Etoh, K.~Takehara, The initial coalescence of
  miscible drops, Phys. Fluids 19~(7) (2007).
\newblock \href {https://doi.org/10.1063/1.2746382}
  {\path{doi:10.1063/1.2746382}}.

\bibitem{yao05}
W.~Yao, H.~Maris, P.~Pennington, G.~Seidel, Coalescence of viscous liquid
  drops, Phys. Rev. E 71~(1) (2005) 016309.
\newblock \href {https://doi.org/10.1103/PhysRevE.71.016309}
  {\path{doi:10.1103/PhysRevE.71.016309}}.

\bibitem{Paulsen2013}
J.~D. Paulsen, Approach and coalescence of liquid drops in air, Phys. Rev. E 88
  (2013) 063010.
\newblock \href {https://doi.org/10.1103/PhysRevE.88.063010}
  {\path{doi:10.1103/PhysRevE.88.063010}}.

\bibitem{paulsen11}
J.~D. Paulsen, J.~C. Burton, S.~R. Nagel, Viscous to inertial crossover in
  liquid drop coalescence, Phys. Rev. Lett. 106~(11) (2011) 114501.
\newblock \href {https://doi.org/10.1103/physrevlett.106.114501}
  {\path{doi:10.1103/physrevlett.106.114501}}.

\bibitem{duchemin03}
L.~Duchemin, J.~Eggers, C.~Josserand, Inviscid coalescence of drops, J. Fluid
  Mech. 487 (2003) 167--178.
\newblock \href {https://doi.org/10.1017/s0022112003004646}
  {\path{doi:10.1017/s0022112003004646}}.

\bibitem{eggers99}
J.~Eggers, J.~R. Lister, H.~A. Stone, Coalescence of liquid drops, J. Fluid
  Mech. 401 (1999) 293--310.
\newblock \href {https://doi.org/10.1017/s002211209900662x}
  {\path{doi:10.1017/s002211209900662x}}.

\bibitem{janssen11}
P.~J.~A. Janssen, P.~D. Anderson, Modeling film drainage and coalescence of
  drops in a viscous fluid, Macromol. Mater. Eng. 296~(3-4) (2011) 238--248.
\newblock \href {https://doi.org/10.1002/mame.201000375}
  {\path{doi:10.1002/mame.201000375}}.

\bibitem{paulsen14}
J.~D. Paulsen, R.~Carmigniani, A.~Kannan, J.~C. Burton, S.~R. Nagel,
  Coalescence of bubbles and drops in an outer fluid, Nat. Commun. 5 (2014)
  3182.
\newblock \href {https://doi.org/10.1038/ncomms4182}
  {\path{doi:10.1038/ncomms4182}}.

\bibitem{aryafar08}
H.~Aryafar, H.~P. Kavehpour, Hydrodynamic instabilities of viscous coalescing
  droplets, Phys. Rev. E 78~(3 Pt 2) (2008) 037302.
\newblock \href {https://doi.org/10.1103/PhysRevE.78.037302}
  {\path{doi:10.1103/PhysRevE.78.037302}}.

\bibitem{jose17}
B.~M. Jose, T.~Cubaud, Role of viscosity coefficients during spreading and
  coalescence of droplets in liquids, Phys. Rev. Fluids 2~(11) (2017) 111601.
\newblock \href {https://doi.org/10.1103/PhysRevFluids.2.111601}
  {\path{doi:10.1103/PhysRevFluids.2.111601}}.

\bibitem{mitra15}
S.~Mitra, S.~K. Mitra, Symmetric drop coalescence on an under-liquid substrate,
  Phys. Rev. E 92~(3) (2015) 033013.
\newblock \href {https://doi.org/10.1103/PhysRevE.92.033013}
  {\path{doi:10.1103/PhysRevE.92.033013}}.

\bibitem{nowak16}
E.~Nowak, N.~M. Kovalchuk, Z.~Che, M.~J.~H. Simmons, Effect of surfactant
  concentration and viscosity of outer phase during the coalescence of a
  surfactant-laden drop with a surfactant-free drop, Colloids Surf. A 505
  (2016) 124--131.
\newblock \href {https://doi.org/10.1016/j.colsurfa.2016.02.016}
  {\path{doi:10.1016/j.colsurfa.2016.02.016}}.

\bibitem{sadeghi18}
H.~M. Sadeghi, B.~Sadri, M.~A. Kazemi, M.~Jafari, Coalescence of charged
  droplets in outer fluids, J. Colloid Interface Sci. 532 (2018) 363--374.
\newblock \href {https://doi.org/10.1016/j.jcis.2018.08.001}
  {\path{doi:10.1016/j.jcis.2018.08.001}}.

\bibitem{chen19}
X.~Chen, P.~Liu, C.~Qi, T.~Wang, Z.~Liu, T.~Kong, Non-coalescence of oppositely
  charged droplets in viscous oils, Appl. Phys. Lett. 115~(2) (2019) 023701.
\newblock \href {https://doi.org/10.1063/1.5109181}
  {\path{doi:10.1063/1.5109181}}.

\bibitem{chen07}
D.~L. Chen, L.~Li, S.~Reyes, D.~N. Adamson, R.~F. Ismagilov, Using three-phase
  flow of immiscible liquids to prevent coalescence of droplets in microfluidic
  channels: criteria to identify the third liquid and validation with protein
  crystallization, Langmuir 23~(4) (2007) 2255--2260.
\newblock \href {https://doi.org/10.1021/la062152z}
  {\path{doi:10.1021/la062152z}}.

\bibitem{crossley10}
S.~Crossley, J.~Faria, M.~Shen, D.~E. Resasco, Solid nanoparticles that
  catalyze biofuel upgrade reactions at the water/oil interface, Science
  327~(5961) (2010) 68--72.
\newblock \href {https://doi.org/10.1126/science.1180769}
  {\path{doi:10.1126/science.1180769}}.

\bibitem{zhang20}
J.-T. Zhang, H.-R. Liu, H.~Ding, Head-on collision of two immiscible droplets
  of different components, Phys. Fluids 32~(8) (2020) 082106.
\newblock \href {https://doi.org/10.1063/5.0018391}
  {\path{doi:10.1063/5.0018391}}.

\bibitem{baumgartner19}
D.~Baumgartner, R.~Bernard, B.~Weigand, G.~Lamanna, G.~Brenn, C.~Planchette,
  Influence of liquid miscibility and wettability on the structures produced by
  drop–jet collisions, J. Fluid Mech. 885 (2019) A23.
\newblock \href {https://doi.org/10.1017/jfm.2019.967}
  {\path{doi:10.1017/jfm.2019.967}}.

\bibitem{bernard20}
R.~Bernard, D.~Baumgartner, G.~Brenn, C.~Planchette, B.~Weigand, G.~Lamanna,
  Miscibility and wettability: how interfacial tension influences droplet
  impact onto thin wall films, J. Fluid Mech. 908 (2020) A36.
\newblock \href {https://doi.org/10.1017/jfm.2020.944}
  {\path{doi:10.1017/jfm.2020.944}}.

\bibitem{xu2022bridge}
H.~Xu, T.~Wang, Z.~Che, Bridge evolution during the coalescence of immiscible
  droplets, J. Colloid Interface Sci. 628 (2022) 869--877.
\newblock \href {https://doi.org/10.1016/j.jcis.2022.08.013}
  {\path{doi:10.1016/j.jcis.2022.08.013}}.

\bibitem{choi14}
S.~B. Choi, J.~S. Lee, Film drainage mechanism between two immiscible droplets,
  Microfluid. Nanofluid. 17~(4) (2014) 675--681.
\newblock \href {https://doi.org/10.1007/s10404-014-1379-x}
  {\path{doi:10.1007/s10404-014-1379-x}}.

\bibitem{liu23}
W.~Liu, J.~M. Park, Numerical study on the engulfing behavior between
  immiscible droplets in a confined shear flow, Chem. Eng. Sci. 266 (2023)
  118265.
\newblock \href {https://doi.org/10.1016/j.ces.2022.118265}
  {\path{doi:10.1016/j.ces.2022.118265}}.

\bibitem{zhang15}
M.-Y. Zhang, H.~Zhao, J.-H. Xu, G.-S. Luo, Controlled coalescence of two
  immiscible droplets for janus emulsions in a microfluidic device, RSC Adv.
  5~(41) (2015) 32768--32774.
\newblock \href {https://doi.org/10.1039/c5ra01718a}
  {\path{doi:10.1039/c5ra01718a}}.

\bibitem{ebadi22}
A.~Ebadi, S.~M. Hosseinalipour, The collision of immiscible droplets in
  three-phase liquid systems: A numerical study using phase-field lattice
  {B}oltzmann method, Chem. Eng. Res. Des. 178 (2022) 289--314.
\newblock \href {https://doi.org/10.1016/j.cherd.2021.12.019}
  {\path{doi:10.1016/j.cherd.2021.12.019}}.

\bibitem{liu22}
W.~Liu, J.~M. Park, Ternary modeling of the interaction between immiscible
  droplets in a confined shear flow, Phys. Rev. Fluids 7~(1) (2022) 013604.
\newblock \href {https://doi.org/10.1103/PhysRevFluids.7.013604}
  {\path{doi:10.1103/PhysRevFluids.7.013604}}.

\bibitem{takamura12}
K.~Takamura, H.~Fischer, N.~R. Morrow, Physical properties of aqueous glycerol
  solutions, J. Pet. Sci. Eng. 98-99 (2012) 50--60.
\newblock \href {https://doi.org/10.1016/j.petrol.2012.09.003}
  {\path{doi:10.1016/j.petrol.2012.09.003}}.

\bibitem{sheely32}
M.~L. Sheely, Glycerol viscosity tables, Industrial \& Engineering Chemistry
  24~(9) (1932) 1060--1064.
\newblock \href {https://doi.org/10.1021/ie50273a022}
  {\path{doi:10.1021/ie50273a022}}.

\bibitem{BERRY2015226}
J.~D. Berry, M.~J. Neeson, R.~R. Dagastine, D.~Y. Chan, R.~F. Tabor,
  Measurement of surface and interfacial tension using pendant drop
  tensiometry, Journal of Colloid and Interface Science 454 (2015) 226--237.
\newblock \href {https://doi.org/https://doi.org/10.1016/j.jcis.2015.05.012}
  {\path{doi:https://doi.org/10.1016/j.jcis.2015.05.012}}.

\bibitem{WACKER20111891}
J.~Wacker, G.~Louis, C.~Razaname, V.~K. Parashar, M.~A. Gijs, Exotic droplets
  formed in microfluidic chips with uniform wettability, Microelectronic
  Engineering 88~(8) (2011) 1891--1893, proceedings of the 36th International
  Conference on Micro- and Nano-Engineering (MNE).
\newblock \href {https://doi.org/https://doi.org/10.1016/j.mee.2010.12.034}
  {\path{doi:https://doi.org/10.1016/j.mee.2010.12.034}}.

\bibitem{Choi2013}
D.~{Choi}, H.~{Lee}, D.~J. {Im}, I.~S. {Kang}, G.~{Lim}, D.~S. {Kim}, K.~H.
  {Kang}, {Spontaneous electrical charging of droplets by conventional
  pipetting}, Scientific Reports 3 (2013) 2037.
\newblock \href {https://doi.org/10.1038/srep02037}
  {\path{doi:10.1038/srep02037}}.

\bibitem{couder05}
Y.~Couder, E.~Fort, C.-H. Gautier, A.~Boudaoud, From bouncing to floating:
  Noncoalescence of drops on a fluid bath, Phys. Rev. Lett. 94~(17) (2005)
  177801.
\newblock \href {https://doi.org/10.1103/PhysRevLett.94.177801}
  {\path{doi:10.1103/PhysRevLett.94.177801}}.

\bibitem{neitzel02}
G.~P. Neitzel, P.~Dell'Aversana, Noncoalescence and nonwetting behavior of
  liquids, Annu. Rev. Fluid Mech. 34~(1) (2002) 267--289.
\newblock \href {https://doi.org/10.1146/annurev.fluid.34.082701.154240}
  {\path{doi:10.1146/annurev.fluid.34.082701.154240}}.

\bibitem{eddi13}
A.~Eddi, K.~Winkels, J.~Snoeijer, Influence of droplet geometry on the
  coalescence of low viscosity drops, Phys. Rev. Lett. 111~(14) (2013) 144502.
\newblock \href {https://doi.org/10.1103/PhysRevLett.111.144502}
  {\path{doi:10.1103/PhysRevLett.111.144502}}.

\bibitem{luo19}
X.~Luo, H.~Yin, J.~Ren, H.~Yan, X.~Huang, D.~Yang, L.~He, Enhanced mixing of
  binary droplets induced by capillary pressure, J. Colloid Interface Sci. 545
  (2019) 35--42.
\newblock \href {https://doi.org/10.1016/j.jcis.2019.03.016}
  {\path{doi:10.1016/j.jcis.2019.03.016}}.

\bibitem{cuttle21}
C.~Cuttle, A.~B. Thompson, D.~Pihler-Puzović, A.~Juel, The engulfment of
  aqueous droplets on perfectly wetting oil layers, J. Fluid Mech. 915 (2021)
  A66.
\newblock \href {https://doi.org/10.1017/jfm.2021.90}
  {\path{doi:10.1017/jfm.2021.90}}.

\bibitem{oratis23}
A.~T. Oratis, V.~Bertin, J.~H. Snoeijer, Coalescence of bubbles in a
  viscoelastic liquid, arXiv preprint arXiv:2305.01363 (2023).
\newblock \href {https://doi.org/10.48550/arXiv.2305.01363}
  {\path{doi:10.48550/arXiv.2305.01363}}.

\bibitem{thoroddsen05}
S.~Thoroddsen, T.~Etoh, K.~Takehara, N.~Ootsuka, On the coalescence speed of
  bubbles, Phys. Fluids 17 (2005) 071703.
\newblock \href {https://doi.org/10.1063/1.1965692}
  {\path{doi:10.1063/1.1965692}}.

\bibitem{hack21}
M.~A. Hack, P.~Vondeling, M.~Cornelissen, D.~Lohse, J.~H. Snoeijer, C.~Diddens,
  T.~Segers, Asymmetric coalescence of two droplets with different surface
  tensions is caused by capillary waves, Phys. Rev. Fluids 6~(10) (2021)
  104002.
\newblock \href {https://doi.org/10.1103/PhysRevFluids.6.104002}
  {\path{doi:10.1103/PhysRevFluids.6.104002}}.

\bibitem{chen17}
X.~Chen, Y.~Sun, C.~Xue, Y.~Yu, G.~Hu, Tunable structures of compound droplets
  formed by collision of immiscible microdroplets, Microfluid. Nanofluid. 21
  (2017) 1--14.
\newblock \href {https://doi.org/10.1007/s10404-017-1944-1}
  {\path{doi:10.1007/s10404-017-1944-1}}.

\bibitem{planchette10}
C.~Planchette, E.~Lorenceau, G.~Brenn, Liquid encapsulation by binary
  collisions of immiscible liquid drops, Colloids Surf. A 365~(1-3) (2010)
  89--94.
\newblock \href {https://doi.org/10.1016/j.colsurfa.2009.12.011}
  {\path{doi:10.1016/j.colsurfa.2009.12.011}}.

\bibitem{yao05viscous}
W.~Yao, H.~J. Maris, P.~Pennington, G.~M. Seidel, Coalescence of viscous liquid
  drops, Phys. Rev. E 71~(1 Pt 2) (2005) 016309.
\newblock \href {https://doi.org/10.1103/PhysRevE.71.016309}
  {\path{doi:10.1103/PhysRevE.71.016309}}.

\bibitem{aarts05}
D.~G. Aarts, H.~N. Lekkerkerker, H.~Guo, G.~H. Wegdam, D.~Bonn, Hydrodynamics
  of droplet coalescence, Phys. Rev. Lett. 95~(16) (2005) 164503.
\newblock \href {https://doi.org/10.1103/PhysRevLett.95.164503}
  {\path{doi:10.1103/PhysRevLett.95.164503}}.

\bibitem{Anthony2017}
C.~R. Anthony, P.~M. Kamat, S.~S. Thete, J.~P. Munro, J.~R. Lister, M.~T.
  Harris, O.~A. Basaran, Scaling laws and dynamics of bubble coalescence, Phys.
  Rev. Fluids 2 (2017) 083601.
\newblock \href {https://doi.org/10.1103/PhysRevFluids.2.083601}
  {\path{doi:10.1103/PhysRevFluids.2.083601}}.

\bibitem{billingham99}
J.~Billingham, Surface-tension-driven flow in fat fluid wedges and cones, J.
  Fluid Mech. 397 (1999) 45--71.
\newblock \href {https://doi.org/10.1017/S0022112099006047}
  {\path{doi:10.1017/S0022112099006047}}.

\bibitem{keller00}
J.~B. Keller, P.~A. Milewski, J.-M. Vanden-Broeck, Merging and wetting driven
  by surface tension, Eur. J. Mech. B Fluids 19~(4) (2000) 491--502.
\newblock \href {https://doi.org/10.1016/S0997-7546(00)00135-7}
  {\path{doi:10.1016/S0997-7546(00)00135-7}}.

\bibitem{keller83}
J.~B. Keller, M.~J. Miksis, Surface tension driven flows, SIAM J. Appl. Math.
  43~(2) (1983) 268--277.
\newblock \href {https://doi.org/10.1137/0143018} {\path{doi:10.1137/0143018}}.

\bibitem{chan11}
D.~Chan, E.~Klaseboer, R.~Manica, Film drainage and coalescence between
  deformable drops and bubbles, Soft Matter 7 (2011) 2235--2264.
\newblock \href {https://doi.org/10.1039/C0SM00812E}
  {\path{doi:10.1039/C0SM00812E}}.

\bibitem{eri10}
A.~Eri, K.~Okumura, Bursting of a thin film in a confined geometry: rimless and
  constant-velocity dewetting, Phys. Rev. E 82~(3 Pt 1) (2010) 030601.
\newblock \href {https://doi.org/10.1103/PhysRevE.82.030601}
  {\path{doi:10.1103/PhysRevE.82.030601}}.

\bibitem{Reyssat06}
E.~Reyssat, D.~Qu\'{e}r\'{e}, Bursting of a fluid film in a viscous
  environment, Europhys. Lett. 76~(2) (2006) 236.
\newblock \href {https://doi.org/10.1209/epl/i2006-10262-x}
  {\path{doi:10.1209/epl/i2006-10262-x}}.

\bibitem{sanjay22}
V.~Sanjay, U.~Sen, P.~Kant, D.~Lohse, {Taylor–Culick} retractions and the
  influence of the surroundings, J. Fluid Mech. 948 (2022) A14.
\newblock \href {https://doi.org/10.1017/jfm.2022.671}
  {\path{doi:10.1017/jfm.2022.671}}.

\bibitem{yokota11}
M.~Yokota, K.~Okumura, Dimensional crossover in the coalescence dynamics of
  viscous drops confined in between two plates, Proc. Natl. Acad. Sci. U.S.A.
  108~(16) (2011) 6395--6398.
\newblock \href {https://doi.org/10.1073/pnas.1017112108}
  {\path{doi:10.1073/pnas.1017112108}}.

\bibitem{baret12}
J.-C. Baret, Surfactants in droplet-based microfluidics, Lab. Chip 12~(3)
  (2012) 422--433.
\newblock \href {https://doi.org/10.1039/C1LC20582J}
  {\path{doi:10.1039/C1LC20582J}}.

\bibitem{beck-broichsitter21}
M.~Beck-Broichsitter, Solvent impact on polymer nanoparticles prepared
  nanoprecipitation, Colloids Surf. A 625 (2021) 126928.
\newblock \href {https://doi.org/10.1016/j.colsurfa.2021.126928}
  {\path{doi:10.1016/j.colsurfa.2021.126928}}.

\bibitem{tian19}
Y.~Tian, W.~Jiao, P.~Liu, S.~Song, Z.~Lu, A.~Hirata, M.~Chen, Fast coalescence
  of metallic glass nanoparticles, Nat. Commun. 10~(1) (2019).
\newblock \href {https://doi.org/10.1038/s41467-019-13054-z}
  {\path{doi:10.1038/s41467-019-13054-z}}.

\end{thebibliography}

\end{document}